\documentclass[lettersize,journal]{IEEEtran}
\usepackage{amsmath,amsfonts}
\usepackage{algorithmic}
\usepackage{algorithm}
\usepackage{array}
\usepackage[caption=false,font=normalsize,labelfont=sf,textfont=sf]{subfig}
\usepackage{textcomp}
\usepackage{hyperref}
\usepackage{stfloats}
\usepackage{pgfplots}
\usepackage{url}
\usepackage{verbatim}
\usepackage{graphicx}
\usepackage{booktabs} 
\usepackage{orcidlink}

\def\BibTeX{{\rm B\kern-.05em{\sc i\kern-.025em b}\kern-.08em
    T\kern-.1667em\lower.7ex\hbox{E}\kern-.125emX}}
\usepackage{balance}
\begin{document}

\title{AI-Hybrid TRNG: Kernel-Based Deep Learning for Near-Uniform Entropy Harvesting from Physical Noise}

\author{Hasan~Yiğit \orcidlink{0000-0002-3832-7055}
        \thanks{H. Yiğit, Muğla Sıtkı Koçman University, Software Engineering Dept.,
        48000 Muğla, Turkey
        (e-mail: \texttt{yigithasan22@gmail.com}; ORCID: 0000-0002-3832-7055).}%

\thanks{}}

\markboth{Journal of \LaTeX\ Class Files,~Vol.~18, No.~9, September~2020}%
{How to Use the IEEEtran \LaTeX \ Templates}

\maketitle

\begin{abstract}

AI-Hybrid TRNG is a deep-learning framework that extracts near-uniform entropy directly from physical noise, eliminating the need for bulky quantum devices or expensive laboratory-grade RF receivers. Instead, it relies on a low-cost, thumb-sized RF front end, plus CPU-timing jitter, for training, and then emits 32-bit high-entropy streams without any quantization step.

Unlike deterministic or trained artificial intelligence random number generators (RNGs), our dynamic inner-outer network couples adaptive natural sources, reseeding, and yields truly unpredictable and autonomous sequences. Generated numbers pass the NIST SP 800-22 battery better than a CPU-based method.  Also, it passes nineteen bespoke statistical tests for both bit- and integer-level analysis. All results satisfy cryptographic standards, while forward- and backward-prediction experiments reveal no exploitable biases.  The model’s footprint is below 0.5 MB, making it deployable on MCUs and FPGA soft cores, as well as suitable for other resource-constrained platforms.

By detaching randomness quality from dedicated hardware, AI-Hybrid TRNG broadens the reach of high-integrity random number generators across secure systems, cryptographic protocols, embedded and edge devices, stochastic simulations, and server applications that need randomness.

\end{abstract}

\begin{IEEEkeywords}
Random Number Generation, Computer Security, Password Generation, Artificial Intelligence, Deep Learning, Neural Networks
\end{IEEEkeywords}

\section{Introduction}

Random numbers are sequences of values with no discernible pattern, forming the backbone of numerous applications in science, engineering, and computer science \cite{Cui, Hellekalek, Deng, Hsieh, Jacak, Bhattacharjee, H.Martín}. Their criticality increases in randomness-dependent domains such as system security, network protection, and artificial intelligence (AI). They are employed for cryptographic key generation, risk modeling in financial analysis, statistical forecasting, test scenario diversification in software engineering, and simulating complex systems \cite{Baldanzi, Oldewurtel, Pham, Z.Hui, J.Chen}.

Traditional random number generation often relies on deterministic algorithms or pre-defined mathematical formulas \cite{Bhattacharjee, KLee}. These pseudorandom number generators (PRNGs) are sensitive to seed selection; if the initial state is predictable, so too is the output sequence \cite{Amirany}. Furthermore, many algorithms suffer from finite cycle lengths, which limit randomness quality over extended usage \cite{M.Matsumoto}. Despite offering flexibility and wide applicability, such techniques are often unsuitable for cryptographic-grade randomness \cite{Corrigan-Gibbs}.

True random number generation (TRNG) techniques, by contrast, harvest entropy from physical phenomena such as thermal noise, quantum behavior, or atmospheric radio signals \cite{KLee, Amirany, Martín}. These methods promise high entropy but typically require dedicated hardware, environmental isolation, and extensive calibration \cite{Mannalatha}. Their lack of portability and high operational cost hinder deployment in edge or embedded platforms.

Compromises in random number quality directly affect cryptographic resilience. Keys exhibiting statistical patterns or low entropy may degrade security—even when using robust algorithms. In block cipher modes, for instance, nonces must be highly random to prevent leakage \cite{Koteshwara}. As threats evolve, the demand for reliable, high-entropy sources becomes more urgent, especially in environments where hardware-based solutions are impractical.

To address these limitations, this study introduces a novel artificial intelligence-based random number generation framework. It eliminates reliance on specialized hardware while supporting reproducibility and scalability. Although initial seeding may incorporate real-world entropy sources, such as RF micro-noise or CPU jitter, the system functions independently at runtime, enabling operation on standard computational platforms.

Our method leverages a kernel-based hybrid deep learning architecture that extracts, enhances, and reseeds entropy through a layered network structure. Deep learning’s powerful representation capabilities are utilized to suppress bias, capture complex distributions, and maintain cryptographic soundness.

This work investigates whether an AI-hybrid TRNG—a truly random number generator that feeds physical noise into a deep-learning model for bias removal and uniformity enhancement—can match or exceed traditional TRNGs in quality, offering a sustainable, portable, and hardware-minimal solution suitable for secure computing and adaptive trusted architectures. The proposed approach integrates signal preprocessing, entropy validation, and neural sequence generation to construct a fully software-defined, high-quality RNG.

\subsection{Literature}
Random numbers have  numerous applications across various fields, including simulation and modeling, statistical analysis, cryptographic systems, quality testing, artificial intelligence, and machine learning. Also, random number generation has two main categories: Pseudo Random Number Generators (PRNGs) and True Random Number Generators (TRNGs) \cite{KLee}. PRNGs operate on deterministic algorithms, producing sequences that appear random but are entirely determined by an initial value known as a seed. Despite their deterministic nature, PRNGs are widely adopted due to their speed and suitability for various applications, including machine learning \cite{Abutaha}. Numerous algorithms exist within this category, each tailored to specific use cases \cite{Deng, Abutaha, Baldanzi}. On the other hand, TRNGs rely on unpredictable physical processes to generate randomness, offering entropy levels higher than PRNGs. These methods leverage the inherent randomness of diverse physical phenomena. While TRNGs provide superior randomness, they may be slower and less practical for specific applications than PRNGs.

The second category comprises True Random Number Generators (TRNGs), which generate random numbers through unpredictable physical processes rather than deterministic algorithms \cite{yu}. In a comprehensive review, researchers \cite{Mannalatha, yu} have further categorized TRNGs into noise, chaos, phase jitter, and other methods. These TRNGs harness the inherent randomness present in physical phenomena to produce sequences of numbers that are truly unpredictable and unbiased. Unlike Pseudo Random Number Generators (PRNGs), which are deterministic and yield repeatable sequences under identical initial conditions, TRNGs offer higher entropy and randomness.

Various studies have proposed innovative TRNG designs leveraging different physical processes. For instance, one study introduces a novel TRNG design utilizing a spin-transfer torque magnetic tunnel junction (MTJ) device \cite{Vatajelu}. Another study explores pseudorandom number generation based on quantum maps, specifically employing a quantum logistic map \cite{Akhshani}. This approach relies solely on equations within quantum chaotic maps, boasting low complexity and minimal hardware requirements, thus enhancing computational speed. Additionally, researchers have developed TRNG circuits based on spin transfer torque magnetic tunnel coupling (STT-MTJ), such as the flexible high speed (RHS)-TRNG circuit, which has been integrated into the RISC-V processor \cite{Fu}. Another proposed TRNG, designed for PVT-tolerant operation, is based on a ring oscillator (RO) with an odd number of inverter stages \cite{Park}. These examples illustrate ongoing research efforts to advance TRNG technology, particularly in hardware-based processes.

In another study \cite{Kumar}, a True Random Number Generator (TRNG) is proposed, employing incident and radio noise-based methods. This approach integrates an acquisition block, front-end circuit, Linear Feedback Shift Register (LFSR), and FPGA to capture and convert atmospheric noise into random numbers. Similarly, in a separate investigation \cite{Lee}, an FM radio signal-based TRNG is introduced. FM radio signals are exploited for their high-entropy characteristics derived from ambient noise, enabling the generation of truly random numbers.

Alternatively, another research \cite{Okada} presents a novel method for random number generation within molecular simulations. Leveraging the inherent randomness in particle coordinates, this approach demonstrates the feasibility of utilizing particle coordinates to generate random numbers. A research \cite{Meiser} proposes a method harnessing the stochastic nature of chemistry by synthesizing DNA strands composed of random nucleotides for random number generation.

Furthermore, another analysis \cite{Li} explores using quantum random number generators (QRNGs) for generating true random numbers on quantum computers. They propose a protocol to enhance the reliability of QRNGs, leveraging the inherent randomness of quantum systems. Reserchers \cite{Cui} suggest utilizing clock jitter from Multi-Stage Frequency Reference Oscillator (MSFRO) for TRNGs, achieving higher throughput and lower hardware resource utilization.

Moreover, in other research \cite{Hsieh} introduces a random number generator based on blockchain and smart contracts. This illustrates the diversification of methodologies in random number generation beyond traditional sources such as circuit and thermal noise.

The research \cite{Oldewurtel} introduces a Stochastic Model Predictive Control (SMPC) framework aimed at addressing climate control within buildings. In the context of this study, previous research has amalgamated Affine Disturbance Feedback with deterministic reconceptualization of chance constraints to manage uncertainties proficiently. Using qualified random sources becomes imperative to faithfully represent the diverse uncertainties inherent in the system. Techniques such as Monte Carlo simulation have been instrumental in scrutinizing the efficacy of control algorithms across varying uncertainty scenarios. 

In streamlining data loading for Random Forests, it’s important to recognize that the algorithm’s built-in randomness—used when selecting variables for each tree—can lead to unstable importance rankings, especially with few trees. Relying on a single run can therefore be misleading \cite{Behnamian}. Incorporating AI-driven feature-selection methods could stabilize these rankings and further reduce data dimensionality.

Similarly, rain-removal techniques that assume rain streaks are temporally random—such as the self-learning model in \cite{WYang}—benefit from genuine randomness in their training data. As AI-generated synthetic datasets become more prevalent (for example, in gaming and vision applications), ensuring high-quality randomness is essential to produce realistic scenarios and maintain model robustness.

The paper \cite{Alvarez} introduces a synthetic dataset for depth estimation and object segmentation tasks within underwater environments. Augmentation techniques are used in both the training and validation subsets to increase the authenticity of the dataset. These techniques include random flips and adjustments to brightness, contrast, and saturation, to align the dataset's characteristics more closely with real-world scenarios. Additionally, grayscale conversion and random resizing of images are implemented. As a result, these augmentation strategies lead to notable enhancements in model performance.

Furthermore, the study underscores the increasing importance of realistic simulation visualizations and the transferability of simulations to real-world scenarios, particularly in fields like robotics. Synthetic datasets, as mentioned, are increasingly utilized in drone competitions \cite{Pham}. Given the challenges associated with creating, processing, and accessing real-world datasets in rapidly evolving domains like robotics, synthetic data creation methods serve as valuable alternatives. The approach proposed in this study contributes to diversifying these datasets, enriching the available resources for advancing research and applications in robotics.

Software testing is a pivotal mechanism in validating whether the actual output aligns with the anticipated production, thus constituting a fundamental component of the software quality assurance process \cite{Z.Hui, J.Chen}. Among the range of testing techniques, random testing (RT) and its refined iterations are paramount in software testing and system reliability \cite{Z.Hui, J.Chen}. However, the efficacy of RT may experience a decrease due to the propensity of failure-inducing inputs to aggregate in specific regions \cite{Z.Hui, J.Chen}. To address this challenge, adaptive random testing (ART) has emerged as a solution, to enhance the diversity of the test suite by evenly disseminating test cases across the entire input space \cite{J.Chen}.

Recent studies have explored the use of deep generative models for pseudo-random number generation. \cite{Okada2023} proposed a learned pseudo-random number generator based on the Wasserstein GAN with Gradient Penalty (WGAN-GP), generating high-entropy bit sequences and validating  via the NIST SP 800-22 test suite. The model demonstrated unpredictability at the bit level where outside its the overfitting area. However, it relies solely on synthetic training data, employs a fixed latent vector as input space, and lacks methodology based on hardware-independent randomness sources.

In parallel, \cite{AN2022} introduced a reinforcement learning (RL)-based framework for random sequence generation, where an agent learns a non-linear policy that maximizes entropy through a custom reward function. The generated sequences were evaluated using entropy scores with the aim of achieving decorrelation, though the study was limited to synthetic data and did not include comprehensive statistical tests such as Chi-Square, ACF, or PSD analysis, nor did it offer comparative validation against natural data sources.

The generation of random test cases relies on a random number generator, for which various pseudorandom number generators (PRNGs) have been documented in the literature. The development of an artificial intelligence-driven random number generator holds promise in furnishing heightened diversity, mainly through its integration into test scenario production processes, thereby mitigating clustering tendencies and augmenting test efficiency.

\section{Method}

This section's methodology for integrating Artificial Intelligence (AI) algorithms into random number generation builds on the foundational information elucidated in the preceding section. The evaluation of results relies on statistics outlined in the same section, designed to assess randomness and scrutinize various facets of the training dataset within the generated dataset. In addition, the goal is to discern disparities through systematic comparisons between training data and generated datasets.

\begin{figure}[!t]
\centering
\includegraphics[width=0.48\textwidth]{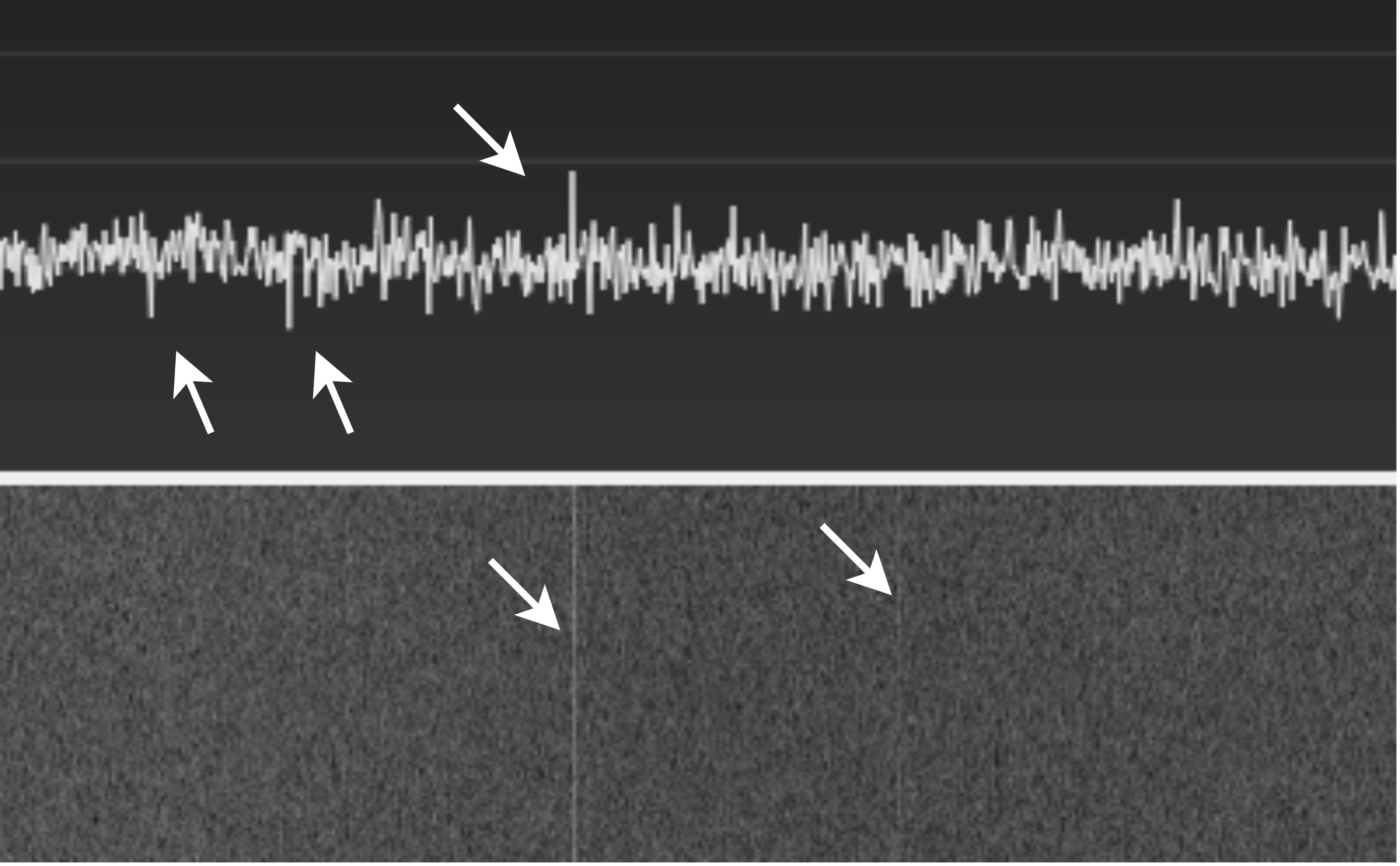}  
\caption{The snapshot displays frequencies with higher power levels than the ambient noise, highlighted by arrows.}
\label{fig:radio-noise2}
\end{figure}

\subsection{Methodology: Kernel-Based Hybrid Layered Learning System}

The theoretical foundation of the proposed kernel-based hybrid layered learning system is grounded in several key concepts from deep learning and neural network theory. The methodology draws upon the strengths of both traditional and contemporary approaches to model training and prediction, aiming to enhance the robustness and accuracy of generated data sequences. Hybrid learning systems combine different learning paradigms to leverage their respective strengths. In this methodology, combining an inner predictive model and an outer adaptive training layer exemplifies the hybrid approach. The inner model operates without weight updates, ensuring prediction stability and consistency, while the outer model dynamically adjusts inputs to optimize performance. Central to the proposed system, Kernel methods facilitate the transformation of input data into a higher-dimensional space where linear separability is more achievable. This transformation enables the model to capture complex relationships within the data, enhancing its predictive capabilities. The use of kernel functions in the outer layer's backpropagation process ensures that input adjustments are informed by these higher-dimensional representations. Layered learning involves a hierarchical approach to model training, where different layers focus on distinct aspects of the learning process. The inner and outer models in the proposed system exemplify this approach, with each layer addressing specific data generation and optimization components. This layered architecture promotes a more structured and efficient learning process.

The proposed methodology introduces a kernel-based hybrid layered learning system. At its core, a conventional deep learning model operates exclusively in a predictive mode, without weight updates. In contrast, the outer layer incorporates a distinct neural network that persistently functions in training mode while generating numerical outputs. The backpropagation algorithm is extended at this stage to adjust input values, with the output of the outer model serving as input parameters for the inner model. Subsequently, the outer layer undertakes a training regimen aimed at minimizing the inner layer’s loss function, leveraging the specialized functionalities of each layer to achieve comprehensive learning.

The architecture comprises flattened and dense layers. The Input Sequence \( S \) is defined as:

\begin{equation} \label{eq:input_sequence}
   S_i = \{s_1, s_2, \ldots, s_{200}\}
\end{equation}

For the outer model, the inputs \( s_i \) are processed to produce adjusted values \( s_i' \).
\begin{equation} \label{eq:outer_layer_adjustment}
s'_{i} = f_{\text{outer}}(S_i), \quad \text{for } i \in \{50, 100, 150, 200\}
\end{equation}

The input \( s_i \) is first subjected to a linear transformation using the weight matrix \( W_{\text{outer}} \) and bias vector \( b_{\text{outer}} \):
\begin{equation}
    S_i''  \leftarrow W_{\text{outer}} \circ  s_i  \quad   \text{for } i \in \{50, 100, 150, 200\}, 
\end{equation}
\begin{equation}
z_i =  f_{\text{inner}}(S_i'')  + b_{\text{outer}}
\end{equation}

where

- \( W_{\text{outer}} \) is the weight matrix of the outer model.

- \( b_{\text{outer}} \) is the bias vector.

The result of the linear transformation \( z_i \) is then passed through a nonlinear activation function \( g \):

\begin{equation}
y_{inner} = g(z_i)
\end{equation}

The outer model is trained to minimize the loss function of the inner model by adjusting the inputs \( S_i'' \). The training objective can be represented as:

\begin{equation}
\min_{W_{\text{outer}}, b_{\text{outer}}} \mathcal{L}_{\text{inner}}(f_{\text{inner}}(S_i''))
\end{equation}

where \( \mathcal{L}_{\text{inner}} \) is the loss function of the inner model. 

The activation function $ g$ is sigmoid employed for the first and last layers in the inner model, while ReLU is utilized for the intermediate layers to introduce nonlinearity. In the gradient descent applied to neural networks, an output of the sigmoid function at 0.5 assumes particular significance within the datasets generated for this research. All sets of randomly generated numeric series in these datasets are uniformly labeled with 0.5. Given the sigmoid function \( \sigma(z) = 0.5 \), we have:

\begin{equation} \label{eq:sigmoid}
\sigma(z) = 0.5 = \frac{1}{1 + e^{-z}}
\end{equation}

Solving for \( z \):

\begin{equation} \label{eq:sigmoid_solution}
-z = 0 \implies z = 0
\end{equation}

This midpoint serves as a decision boundary, highlighting the model's indifference between the two classes. This ensures that no input number set is distinctly associated with any class. Consequently, the desired accuracy is set to 0 for all training, validation, and test datasets in training the proposed kernel deep learning model. This decision aligns with the understanding that an output of 0.5 corresponds to a state of uncertainty, and the model should refrain from confidently assigning data points to specific classes.

During the initial training phase (before switching to prediction mode), the weights and biases are optimized to minimize a loss function, typically Mean Absolute Error (MAE) in this context:

\begin{equation}
\mathcal{L}_{\text{inner}} = \frac{1}{n} \sum_{i=1}^{n} |y_i - \hat{y}_i|
\end{equation}

where \( y_i \) is the true value and \( \hat{y}_i \) is the predicted value from the inner model. At this point, \( y_i = 0.5 \) and \( \hat{y}_i = y_{\text{inner}} \).

The parameters \( W_{\text{outer}} \) and \( b_{\text{outer}} \) which are calculated according to Equation \ref{ex4} are updated using backpropagation based on the gradient of the loss function with respect to these parameters :

\begin{equation}
W_{\text{outer}}'\leftarrow W_{\text{outer}} - \eta \frac{\partial \mathcal{L}_{\text{inner}}}{\partial W_{\text{outer}}}
\end{equation}

\begin{equation}
b_{\text{outer}}' \leftarrow b_{\text{outer}} - \eta \frac{\partial \mathcal{L}_{\text{inner}}}{\partial b_{\text{outer}}}
\end{equation}

where \( \eta \) is the learning rate.
Then, the weight update operates according to the given equations. The trained outer model weights will not be used. Thus,  \( s_{i} \) must be updated to prevent the outer model results. To achieve this, \( w' \) values in the first layers are set to 1:

\begin{equation} \label{eq:update_weights}
s_{i}W_{\text{outer}}' = y
\end{equation}

Set \( W_{\text{outer}}' \) to 1:

\begin{align} \label{eq:update_equations}
s_{i}'1 &= y' \\
y' &= y \\
s_{i} W_{\text{outer}}' &= s_{i}'1
\end{align}

Thus, we get updated values by using \( f \), which is update function, in our defined learning system for shifting time \( t_0 \):

\begin{equation} \label{eq:updated_values}
\{s_{50}', s_{100}', s_{150}', s_{200}'\} = f(\{s_{50}, s_{100}, s_{150}, s_{200}\})
\end{equation}

Then, the left shifting operation is applied to the sequence \( S \). After generating \( s_{50}', s_{100}', s_{150}', s_{200}' \), the sequence \( S \) is shifted left by 1 position:

\begin{equation} \label{eq:left_shift}
s_{i-1} = s_{i}, \quad \text{for } i \in \{50, 100, 150, 200\}
\end{equation}

We get a completely new sequence after 50 shifting operations:

\begin{equation} \label{eq:new_sequence}
\begin{split}
S' = \{s_1, s_2, \ldots, s_{49}, s_{50}', s_{51}, \ldots, s_{99}, s_{100}', s_{101}, \\ \ldots, s_{149}, s_{150}', s_{151}, \ldots, s_{199}, s_{200}'\}
\end{split}
\end{equation}

In sequence \( S' \), \( s_1 \) equals \( s_{50-t_0}' \) and \( s_2 \) equals \( s_{50-t_1}' \). If we get the values according to the time \( t \), \( s_{i} \) is placed at 50 shifting time intervals and created simultaneously on the time axis. The inner model takes all of \( S \) in every operation, so \( s_{i} \) is affected by all values. We believe that these two methods contribute to supporting continuous and robust random number set generation.

\begin{table*}[ht]
\centering
\caption{Comparison of Shannon and min-entropy per float between raw physical mantissa data and AI-generated outputs. The training set consists of 5,000 files, the AI-generated output comprises 44,000 files, and the seed dataset contains 44,999 files.}
\begin{tabular}{l>{\centering\arraybackslash}p{2cm}>{\centering\arraybackslash}p{2cm}>{\centering\arraybackslash}p{2cm}>{\centering\arraybackslash}p{2cm}}
\toprule
\textbf{Dataset} & \textbf{Avg. Shannon Entropy} & \textbf{Avg. Min-Entropy} & \textbf{Shannon--Min Gap} & \textbf{$p_\text{max}$} \\
\midrule
Raw Training (*.pkl) & $7.617 \pm 0.016$ & $6.697 \pm 0.273$ & $0.920$ bits & $\approx 1/104$ \\
Raw Seed (*.pkl) & $7.617 \pm 0.016$ & $6.696 \pm 0.269$& $0.921$ bits& $\approx 1/104$ \\
AI Output (*.pkl)   & $7.644 \pm 0.0001$& $7.641 \pm 0.052$& $0.0003$ bits& $\approx 1/200$ \\
\bottomrule
\end{tabular}
\label{tab:entropy_comparison}
\end{table*}

\subsection{Data Set Generation}

In this research, the evaluation of the proposed approach involved the generation of two distinct datasets to assess its efficacy. The initial dataset was meticulously crafted using the secrets module CPU-based generation of the random numbers from operating system-specific entropy \cite{secrets}. On the other hand, the second dataset, derived from the recorded radiofrequency (RF) noise, presented more complex challenges. The data collection process required meticulous attention to detail due to inherent challenges. In particular, the extensive utilization of frequency bands posed the risk of introducing patterns into the recorded dataset, contrary to the objective of achieving a patternless dataset. Figure \ref{fig:radio-noise2} illustrates a scenario where many different frequencies exhibit higher power levels than ambient noise, potentially introducing patterns. The waterfall representation in the corresponding figure reveals apparent patterns along the time axes. Another challenge arises from the limited frequency gap between broadcast signals. To address this, dataset recording focused on frequency gaps, necessitating careful consideration of local oscillator frequencies.

Local oscillator frequencies, also known as carrier frequencies, pose a challenge due to their natural repeating patterns resulting from mixer circuits. This has led to the need for strategic solutions to eliminate unwanted patterns in atmospheric noise recordings. It was necessary to overcome these challenges in the dataset by providing a real-world example of this phenomenon in Figure \ref{fig:radio-noise2}. Following experiments addressing these mentioned problems, the RF noise data set was meticulously recorded in a single continuous segment lasting more than 90 minutes using frequency modulation techniques. The recorded data set was recorded in lossless waveform format (wav) with  resampled. This dataset, consisting of more than 43 million data points, served as the primary source for further research and analysis \cite{Dataset}. This step ensured consistency in the network training procedure. The comprehensive approach to dataset generation established a robust foundation for evaluating the proposed methodology, achieving a delicate balance between high randomness and patternless characteristics in RF noise recordings.

\begin{figure*}[t]
  \centering
  \includegraphics[width=\linewidth]{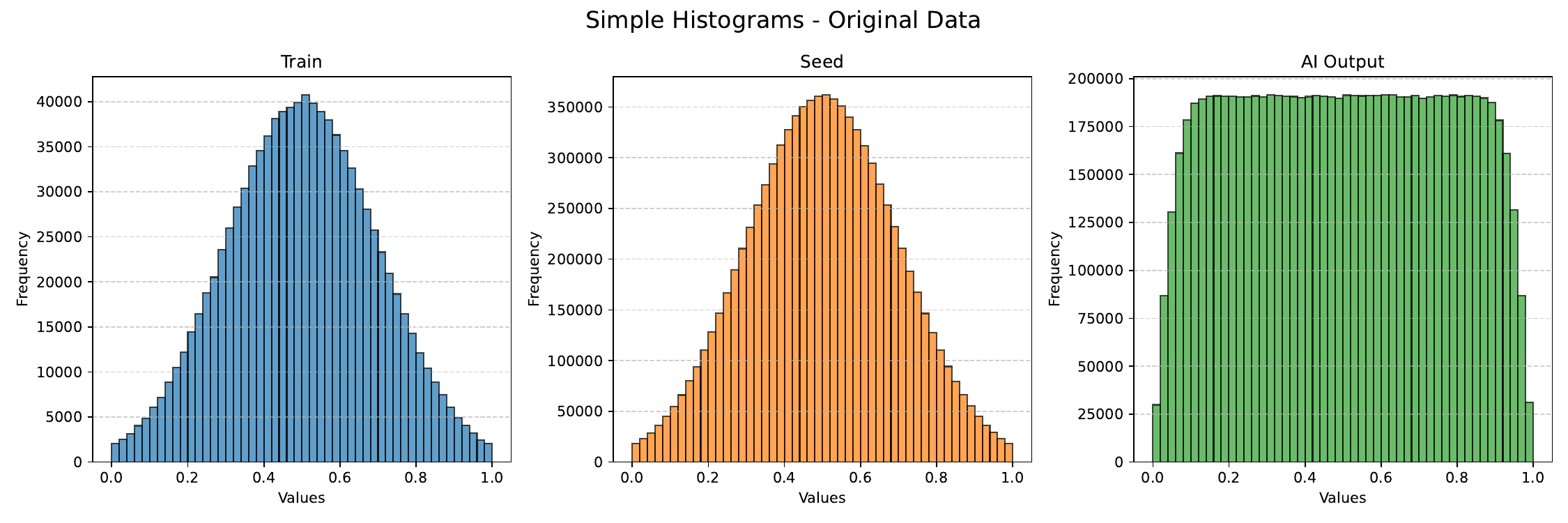}
  \caption{Empirical distribution of original float values
           extracted from the raw training data (\emph{left}),
           the physical seed block (\emph{middle}),
           and the AI‐generated output stream (\emph{right}).
           The raw and seed histograms retain a Gaussian-like peak,
           whereas the AI output is almost perfectly flat over $[0,1)$,
           confirming that the kernel-based extractor removes amplitude bias
           and produces a uniform symbol distribution before further hashing.}
  \label{fig:hist}
\end{figure*}

\section{Test Results} \label{Test-Results}

This section evaluates the effectiveness of the proposed kernel-based hybrid layered AI algorithm in generating statistically robust random sequences. To assess the randomness quality, the model was rigorously tested using two distinct datasets and a broad range of statistical methods. These tests examine structural, distributional, and entropy-related characteristics, aiming to validate the system’s performance compared to original datasets and existing methods.

A comprehensive set of statistical tests was conducted, as summarized in Table \ref{table:pass-rates}. These include stationarity tests such as the Augmented Dickey-Fuller (ADF) and Phillips-Perron, autocorrelation analysis (ACF), entropy estimation, Chi-square and Kolmogorov-Smirnov tests for distributional alignment, spectral density analysis, and goodness-of-fit evaluations (predefined distrubitions check) using Weibull, Gamma, Lognormal, and Poisson distributions and runs test for randomness. The combination of time-domain and frequency-domain metrics provides a multidimensional assessment of the generated data.

The results indicate that the AI-generated sequences exhibit superior statistical properties compared to the original datasets. While the original radio data pass ratio is 97.2\% , AI-trained model (on radio datasets) pass ratio 100\% without distrubition check tests and Spearman-Kendall Tau's tests. Additionally, thers is no failfure on distrubitions test means randomnes dos not fit any distrubition. Only two tests failed Spearman-Kendall Tau's test. Notably, the proposed model effectively replicates key characteristics such as entropy levels and autocorrelation patterns and all off them passed.

\begin{table}[ht]
\centering
\caption{NIST SP 800-22 randomness test results 100 x 1Mb,  *.bin files. }
\begin{tabular}{p{3.5cm} p{1cm} p{1cm} p{1cm}}
\toprule
\textbf{Test} & \textbf{Sub-test} & \textbf{Pass Rate (\%) Ours}&\textbf{Pass Rate (\%) CPU}\\
\midrule
Monobit & -- & 100  &100.0\\
Frequency Within Block & -- & 100.0  &97.0\\
Cumulative Sums     & Forward & 99.0  &100.0\\
                    & Backward & 98.0  &100.0\\
Runs & -- & 99  &99.0\\
Longest Run of 1's in a Block & -- & 99.0  &97.0\\
Binary Matrix Rank & -- & 99.0  &99.0\\

Discrete Fourier Transform & -- & 99.0  &98.0\\
Non-Overlapping Template Matching & sub 148 & 99.0  &99.1\\
Overlapping Template Matching & --& 100.0  &100.0\\
Maurer’s Universal & -- & 100.0  &98.0\\
Approximate Entropy & -- & 100.0  &99.0\\
Serial & $m-1$ & 99.0  &97.0\\
       & $m-2$ & 100.0  &99.0\\
Linear Complexity & -- & 99.0  &100\\

Random Excursion & $x \in [-4, +4]$ & $\approx$62.5  &60.5\\
Random Excursion Variant & $x \in [-9, +9]$ & $\approx$62.5  &60.0\\

\midrule

\bottomrule
\end{tabular}
\label{tab:NISTResults}
\end{table}

Visual analyses further support these findings. As shown in Figure \ref{fig:re3}, the runs test statistics in the AI-generated data are more symmetrically distributed around zero and all p values are bigger than 0.01, while entropy curves closely resemble those of the original dataset (around 7.62)—sometimes with improved consistency. Figure \ref{fig:re2} shows that the average lag correlation for all test around 0 means series behaves like random in the generated data clusters tightly around zero, suggesting reduced temporal dependence.  Some tested series in some specific lag values reaches 0.25 values means there are small fluctuations but they are not statistically significant. Also all power at 0 Hz means series behaves random.  The Kendall Tau and Spearman correlation analyses, supported by the generated-to-train data ratio differences presented in Table \ref{table:pass-rates} and \ref{tab:NISTResults} and illustrated in Figure \ref{fig:re1}, confirm that there is no statistically significant correlation between the generated and trained data of the proposed AI model. Additionally, Several authors have \cite{Okada2023} prove artificial intelligence can create randomness even if use  poor random input seeds.

To assess the statistical quality of the generated bitstreams, we applied the full NIST SP 800-22 randomness test suite used from \cite{NistTest} to binary datasets constructed from \texttt{*.pkl} files, with each 1 MiB stream produced by seeding a SHA-512–based Hash\_DRBG with 4 600 raw mantissa bits (extracted from 200 single-precision floats per pickle) and evaluating all 100 resulting streams independently across the suite’s 188 sub-tests.

\begin{table}[ht]
\centering
\caption{Cryptographic robustness test results over 409,600 blocks (each 2048 bits). Metrics include Hamming distance uniformity, autocorrelation, and next-bit predictability via logistic regression.}
\begin{tabular}{>{\raggedright\arraybackslash}p{2.3cm} 
                >{\centering\arraybackslash}p{1cm} 
                >{\centering\arraybackslash}p{1cm} 
                >{\centering\arraybackslash}p{1cm} 
                >{\centering\arraybackslash}p{1.5cm}}
\toprule
\textbf{Test Metric} & \textbf{Failure Rate} & \textbf{Fail condition} & \textbf{Mean} & \textbf{Min / Max} \\
\midrule
Hamming Distance (p-value)     & 0.94\%& $< 0.01$ & 0.518& $0.000 / 1$\\
Hamming Distance (raw bits)    & --     & --& $512 \pm 22.6$& 440 / 586\\
Autocorrelation (ACF Max)      & 2.56\%& $> 0.10$  & 0.067 & 0.024 / 0.164\\
Forward Accuracy (Next-Bit)    & 0.19\%& $> 0.6$  & 0.500 & 0.332 / 0.668\\
Backward Accuracy (Next-Bit)   & 0.19\%& $> 0.6$  & 0.500 & 0.337 / 0.673\\

\bottomrule
\end{tabular}
\label{tab:crypto_tests}
\end{table}

As shown in Table~\ref{tab:NISTResults}, the proposed generation pipeline passed all major tests. Across the 100 × 1 MiB generated binaries, virtually every NIST SP 800-22 sub-test shows excellent uniformity: Monobit, Frequency-Within-Block, Runs, DFT, matrix-rank, Maurer’s Universal, Approximate Entropy, Serial, and Linear Complexity all exceed a 99 \% pass rate, with several hitting a perfect 100 \%. Both directions of the Cumulative Sums test remain above 98 \%, and the Non-Overlapping Template test (148 templates) also clears 99 \%. The only area that still looks weak is the Random Excursions (RE) and Random Excursions Variant (REV) suite, where barely 62.5 - 63 \% of the state-specific sub-tests record a “pass,”.  Most of those apparent “failures,” however, stem from sequences in which the random walk visits a given state fewer than J = 500 times—the NIST threshold below which the sub-test is not evaluated. In other words, the data are simply too short (or too well balanced) for many RE/REV sub-tests to run, and the harness counts each “not run” as a miss. We confirmed the same behavior with bitstreams generated by CPU-based —widely regarded as cryptographically secure—which scored virtually the same 60.5\% and 60\% on RE/REV under identical conditions.

As shown in Table~\ref{tab:entropy_comparison}, the proposed generator achieves a near-uniform entropy transfer from the 200-float physical seed to the output space. The average Shannon entropy increases slightly from $7.617$ to $7.644$ bits per float. More importantly, the min-entropy improves substantially, rising from $6.696$ to $7.641$ bits. This reduction in the Shannon--min gap from $0.92$ to just $0.0003$ bits indicates that the AI extractor not only preserves entropy, but also \emph{eliminates statistical bias}. The maximum symbol probability ($p_\text{max}$) drops from approximately $1/104$ in the raw mantissa to $1/200$ in the AI output, further confirming the whitening effect. Although no new entropy is created, the generator makes full use of the available randomness from physical noise, effectively redistributing it across the output symbol space. 

The cryptographic robustness of the generated sequences was evaluated using multiple statistical tests over 409600 non-overlapping 2048-bit blocks.  Only 0.94\% of blocks failed the Hamming distance independence test ($p < 0.01$), while autocorrelation violations ($> 0.1$) occurred in 2.56\% of cases,  well within acceptable bounds for secure generation.  Next-bit predictability was assessed using logistic regression accuracy on 16-bit sliding windows. Only 0.19\% (forward) and 0.19\% (backward) of blocks exceeded the predictability threshold of 0.6, with mean accuracies near 0.500 — suggesting no exploitable patterns.  These results indicate strong forward secrecy and statistical unpredictability, reinforcing the generator’s robustness in practical cryptographic applications.

These findings demonstrate that the generator's output maintains high statistical independence, resists simple predictive attacks, and exhibits uniformity both in bit-level distribution and in inter-block structure. This reinforces its suitability as a source of cryptographic randomness under constrained or adversarial conditions.

In summary, the proposed algorithm produces statistically sound and cryptographically resilient random sequences. The combination of successful standard randomness tests, low predictability, entropy stability, and adaptability across datasets validates its potential for use in critical applications requiring high-quality randomness.

\begin{table*}[ht]
\caption{Comparison of AI-Hybrid TRNG with Prior AI-based RNGs}
\label{tab:literature_comparison}
\centering
\begin{tabular}{p{3.5cm} p{4.1cm} p{4.1cm} p{4.1cm}}
\hline
\textbf{Criteria} & \textbf{AI-Hybrid TRNG (Ours)} & \textbf{Upside-Down RL PRNG~\cite{AN2022}} & \textbf{WGAN-GP LPRNG~\cite{Okada2023}} \\
\hline

Model Architecture & Kernel-based hybrid, frozen inner + adaptive outer reseed & RL agent with entropy-reward, online permutation & GAN with WGAN-GP structure \\

Physical Data Source & True entropy from RF and CPU jitter & None; fully synthetic & None; uses MT19937 \\

True vs. Pseudo Origin & True entropy& Pseudo& Pseudo\\

Output Type & Native 32-bit float, $\sim$10ms/512-bit block & 128-bit block output & Float $[0,1)$, bit-sliced \\

Entropy Budget & 200-float seed $\rightarrow$ 1.52kbit& Limited by seed and weights & MT seed only \\

Test Coverage & NIST SP 800-22, Shannon/Min-Entropy, ACF, PSD, etc. (19 total) & NIST SP 800-22 only & NIST SP 800-22 only \\

Test Results &  Avg. NIST SP 800-22 without RE \& REV 99.3\%, RE \& REV 62.5\% $H_\infty$ loss $<$ 0.03 bit& 5 Test $<$50\%,  RE 56\% REV 33\%& NIST SP 800-22 is good but; fails after 450k iters \\

Forward / Backward Security & Hamming dist. $\approx$ ideal; next-bit $\approx$ 0.5 & Evaluated HD and next-bit & Not analyzed \\

Overfitting & No overfit; outer retrains per cycle & Online RL stabilizes & Overfits after 450k iters \\

Hardware / Portability & SDR only for seed; pure firmware at runtime & Fully software & Fully software \\

Strengths & Physical noise, high entropy, 19 tests passed, portable & Formal RL framework, adaptive reward & Novel GAN design \\

Weaknesses & Needs  seed from physical source& No true entropy, seed exposure risk & Entropy drop, no forward secrecy \\

\hline
\end{tabular}
\end{table*}

\section{Discussion}

The experimental evidence presented in Section \ref{Test-Results} establishes three firm facts about the proposed kernel-based generator. First, it preserves virtually all of the physical entropy contained in the 200-float seed.  The Shannon is $7.644 \pm 0.0001$  and the Shannon–min gap shrinks to $0.0003$ bits (Table \ref{tab:entropy_comparison}); within measurement.
Second, the generator removes bias instead of masking it: the most-probable mantissa symbol falls from 1/104 in the raw data to 1/200 after extraction, yielding a empirical flat histogram in Figure \ref{fig:hist}.
Third, All 19 high-order diagnostics including PSD, $X^2$, KS, ADF, ACF, and rank-correlation tests operate on the float-domain sequence after a deterministic ×1000 scaling. Working in the[0,1000) integer space preserves roughly 10 bits of resolution per sample and exposes spectral or monotonic artefacts that would be invisible in bit-level data. Pass rates above 97 \% across these tests confirm that the generator’s raw numeric output is free of hidden periodicity, drift, or nonlinear correlation before any hash whitening is applied.
Fourth, the output passes stringent randomness diagnostics. The NIST SP 800-22 battery reports a 99.3\% average pass-rate in Table \ref{tab:NISTResults}  without RE and REV test. These test results explained in Section \ref{Test-Results}.  Another  block-level tests of 409 600 × 2048-bit segments show lower 1\% Hamming-p failures, $<$ 2.6\% autocorrelation excesses, and $\approx$0.2\% next-bit–prediction failures (Table \ref{tab:crypto_tests}). Together these results confirm that the bit-stream is statistically independent, unpredictable, and free of short- or long-range structure.

In practical terms the architecture is small, fast, and hardware-light.  Optimization reduces the model footprint from five megabytes to just 0.5 MB, small enough for commodity micro-controllers.  Kernel operations finish in 10–11 ms on a consumer GPU and the full pipeline delivers 200 high-entropy floats in under one second on a Ryzen 9 7900X CPU, 32GB RAM, and RTX 3070Ti GPU.  All heavy analogue hardware is confined to the seeding phase—an inexpensive SDR captures RF micro-noise and CPU jitter—so the deployed firmware needs no extra circuitry at run-time.

A qualitative comparison with recent Learned PRNGs clarifies where the contribution sits in the literature.  GAN-based LPRNG \cite{Okada2023} and Upside-Down RL PRNG \cite{AN2022} rely on synthetic latent vectors or MT19937 samples; their entropy budget never exceeds the private seed, and statistical bias is removed only after an external hash.  By contrast our generator (i) starts from real-world entropy, (ii) proves near-lossless transfer inside the network itself, and (iii) compresses the entire model into half a megabyte, making it deployable on MCUs and FPGAs.  Table \ref{tab:literature_comparison} summarises these architectural and performance gaps.

There are, nonetheless, clear boundaries to the claim.  The system does not manufacture new entropy—the level of cryptographic secrecy still depends on regular reseeding from physical noise (we assume that there is on system entropy leackage form the hardware).  

Overall, our results demonstrate that a compact AI-hybrid TRNG extractor can transform a modest, hardware-sourced entropy seed into a statistically robust random bit stream, thereby avoiding the bulk and complexity of full-fledged TRNG hardware. By leveraging a minimal physical seed for initialization and then relying on AI-driven bias removal and uniformity enhancement, the proposed generator preserves entropy, eliminates bias, and imposes minimal computational overhead. This makes it an ideal randomness core across a broad spectrum of modern systems—from high-performance computing clusters and enterprise servers to edge devices, IoT sensors, and embedded architectures—where portable, high-quality entropy sources with only minimal seed-hardware requirements are critically needed.

\section{Conclusion}

We have introduced a \emph{kernel-based AI‐hybrid TRNG} that combines lightweight physical noise sources (RF micro-noise and CPU jitter) with a deep-learning extractor.

Comprehensive evaluation shows that the system

\begin{itemize}
    \item Near-uniform entropy preservation: 
The mantissa pipeline loses less than $0.003$ bit per symbol,
while raising the min-entropy from $\sim\!6.7$ to $\sim\!7.6$ bits/symbol,
corresponding to a drop in the maximum symbol probability
$p_{\mathrm{max}}$ from~$\approx\!1/104$ to~$\approx\!1/200$.

  \item Full-entropy output after conditioning:
        When a Hash-DRBG whitener is applied, the resulting byte stream
        achieves an \emph{average} Shannon entropy of
        $\approx7.9998$ bits / byte and a \emph{worst-case}
        min-entropy of $\approx7.9376$ bits / byte, averaged over
        $100$ binary files of $\sim1$ MiB each.
  \item retains a formally bounded physical entropy budget, the generator consistently passes NIST randomness tests (only condition future work will focus on RE and REV tests) as well as forward- and backward-secrecy evaluations;
\item requires only a 0.5\,MB firmware footprint on commodity MCUs or FPGAs—no ring-oscillator arrays or dedicated TRNG blocks—thereby enabling seamless deployment across IoT and edge platforms;

\end{itemize}

These results demonstrate that AI-hybrid TRNG extraction can preserve and whiten low-cost physical entropy sources with negligible statistical loss, offering a viable trust anchor for adaptive-security architectures, cryptographic key generation, and large-scale simulations.

Future work will (i) quantify long-range correlation with
TestU01 and the SP 800-90B IID/non-IID entropy estimators;
(ii) diagnose and improve Random Excursions (RE) and
Random Excursions Variant (REV) performance—for instance
by increasing walk length; (iii) explore alternative kernel
topologies that curb expansion-induced dilution; and
(iv) investigate seamless reseeding from heterogeneous
sensors (e.g., Bluetooth RSSI or on-die thermal noise).
By tightening these bounds, we aim to establish a practical,
formally verified, and fully portable pathway for AI-enhanced
true-entropy generators in safety-critical systems.

\section*{Supplementary Material}

The supplementary material includes three parts:

\begin{enumerate}
    \item \textbf{Background \& Pre-deployment Tests:} Mathematical foundations of the kernel-based entropy extractor, test details, plus pre-deployment analyses results.
    \item \textbf{Experimental Setup:} Experimental Setup: Images of the SDR-based entropy acquisition system and details about the hardware and recording parameters, including CPU and GPU usage metrics. The proposed model architecture includes the direct model code block.
    \item \textbf{Statistical Test Results:} Complete NIST SP 800-22 logs for both the CPU baseline and the proposed AI-hybrid TRNG. Test data publicly available on Zenodo \cite{yigit2025testdata}.

    \item \textbf{Binary Datasets:} Two corpora of 100 × 1 MiB raw bitstreams—Group A (AI-hybrid TRNG) and Group B (CPU baseline).

\end{enumerate}

        %


\clearpage 
\onecolumn          

\section*{Supplementary Material- Background  Theory\& Pre-deployment Tests}

\setcounter{page}{1}

\setcounter{figure}{0}
\renewcommand{\thefigure}{S\arabic{figure}}

\setcounter{table}{0}
\renewcommand{\thetable}{S\arabic{table}}

\setcounter{equation}{0}
\renewcommand{\theequation}{S\arabic{equation}}

\label{sec:supp-background}

In this research, we develop novel kernel-based neural structures for creating AI-based random number series. We explore artificial neural networks and the backpropagation algorithm, which are fundamental to our approach. The research includes a theoretical backdrop covering function conditionals and iterative constructs, focusing on a designed backpropagation algorithm rooted in calculus principles. Additionally, we enhance clarity by conducting statistical analyses to assess the capability of generating random number sequences under various tests, enriching the interpretative framework applied to the results.

The perceptron, a core unit in artificial neural networks, processes numerical input values by multiplying them with corresponding weights and applying an activation function for non-linearity, enabling it to capture intricate patterns in the data for neural networks \cite{Goodfellow}. Error calculation and backpropagation are crucial for refining its performance. The error is computed by comparing the output to the expected output, and the gradient of the error with respect to the weights is used to adjust them, enhancing the perceptron's learning capacity.

\subsubsection{Forward Propagation }

In neural networks, forward propagation demonstrates The Equation \ref{ex1} involves computing the weighted sum of input features and applying an activation function. The output \(y\) is mathematically defined as the activation function \(f\) applied to the dot product of weights and inputs, along with a bias term \(b\)  \cite{Goodfellow}:

\begin{equation}
\label{ex1}
 y = f\left(\sum_{i=1}^{n} (w_i \cdot x_i) + b\right)
\end{equation}

where \(y\) signifies the network's output, \(f\) represents the activation function, \(w_i\) and \(x_i\) denote weights and inputs, the sum is over all input features (\(i = 1, 2, 3\)), and \(b\) stands for the bias term \cite{Goodfellow}.

\subsubsection{Back Propagation}

Before backpropagation in a neural network, error computation involves selecting a loss function to quantify the difference between the predicted output (\(y_{\text{predicted}}\)) and the actual target output (\(y_{\text{actual}}\)). This error formulation serves as the foundation for subsequent optimization steps \cite{Goodfellow}. Various loss functions may be considered based on the nature of the task during the loss function selection process.

In neural network optimization, weight updates minimize errors during training. Activation functions enhance this process, working synergistically with the network architecture. Backpropagation, rooted in calculus principles, is central to this dynamic optimization framework. At its core is the chain rule, which facilitates derivative calculations for composite functions—essential within neural networks. Equation \ref{ex4} succinctly represents the chain rule's essence \cite{Goodfellow}.

\begin{equation}
\label{ex4}
\frac{d(f(g(x)))}{dx} = \frac{df}{dg} \cdot \frac{dg}{dx}
\end{equation}

This concise representation proves instrumental in discerning the cascading effects of parameter modifications in one layer upon the ultimate output or loss. The weight update method anchors the optimization process, with the comprehensive weight update expression by \ref{ex5} taking the form\cite{Goodfellow}:

\begin{equation}
\label{ex5}
w' = w - \eta \cdot \frac{\partial E}{\partial w}
\end{equation}

where \(w'\) signifies the updated weight, \(w\) is the current weight, \(\eta\) (eta) represents the learning rate, and \(\frac{\partial E}{\partial w}\) denotes the partial derivative of the error (loss) concerning the weight \(w\). Delving into the mathematical foundations  \cite{Goodfellow} unveils a granular examination of the weight update formula \ref{ex6}:

\begin{equation}
\label{ex6}
w' = w - \eta \cdot \frac{\partial E}{\partial \sigma} \cdot \frac{\partial \sigma}{\partial z} \cdot \frac{\partial z}{\partial w}
\end{equation}

Here, \(\frac{\partial E}{\partial \sigma}\) denotes the partial derivative of the error concerning the output activation function \(\sigma\), \(\frac{\partial \sigma}{\partial z}\) captures the partial derivative of the activation function, and \(\frac{\partial z}{\partial w}\) reflects the partial derivative of the weighted sum concerning the weight \(w\)—notably, \(\frac{\partial z}{\partial w_i} = x_i\) due to the inherent nature of the weighted sum in a perceptron.

\subsection{Statistical analysis of random number series}

This section explores some statistical analyses applied to random number series to comprehensively understand the study output related to random number series defined in the scope of these tests, paving the way for informed decision-making in various domains. By systematically exploring statistical techniques, we aim to enhance comprehension of the study's output related to random number series, enabling informed decision-making across domains.

\subsubsection{Augmented Dickey-Fuller (ADF) Test}
\cite{ADF, Fuller} is a statistical method used to determine the stationarity of a time series by assessing the unit root property. It distinguishes between stationary and nonstationary dynamics within the series by examining a regression model that incorporates lagged differences. The test evaluates the unit root hypothesis, where a non-zero coefficient implies nonstationarity. Facilitating the execution of such statistical evaluations, the statsmodels \cite{statsmodels} library offers a robust framework. The foundational formulation of the ADF test is expressed through the following regression model \ref{ex7}:

\begin{equation}
\label{ex7}
\begin{split}
\Delta y_t &= \alpha + \beta t + \gamma y_{t-1} + \\
&\quad \delta_1 \Delta y_{t-1} + \delta_2 \Delta y_{t-2} + \ldots + \delta_{p-1} \Delta y_{t-(p-1)} + \varepsilon_t
\end{split}
\end{equation}

Here, \(y_t\) signifies the time series, \(\Delta y_t\) represents the first difference of the series (\(y_t - y_{t-1}\)), \(t\) denotes the trend over time, \(\alpha\) is the constant term, \(\beta\) is the trend coefficient, \(\gamma\) is the Lag 1 coefficient, \(\delta_1, \delta_2, \ldots, \delta_{p-1}\) are coefficients of lagged differences, and \(\varepsilon_t\) represents the error term  \cite{ADF, Fuller}.

Interpreting the ADF test results relies on the p-value: a value below 0.05 or 0.01 indicates stationarity, while higher values suggest potential nonstationarity. Rejecting the null hypothesis (\(H_0: \gamma = 0\)) implies stationarity. It's important to note that stationarity in a time series means consistent statistical properties over time beyond a specific numerical range. In random sets, a stationary time series is generally expected.

\subsubsection{The Runs Tests}

 assesses patterns and deviations from randomness, offering a nonparametric alternative when distribution assumptions are uncertain. It computes various tests for numeric, binary, or categorical data. For numeric data, it calculates the Expected Number of Runs (\(E(R)\)) defined as equation \ref{ex8} and the Test Statistic (\(Z\)) defined statement \ref{ex9} based on the number of occurrences of distinct values in the sequence. For numeric data, exact and asymptotic Wald-Wolfowitz Runs Tests for Randomness are computed, considering the number of runs above and below a designated reference value \cite{Gibbons, Rukhin, Karras}. 

\begin{equation}
\label{ex8}
E(R) = \frac{2n_1n_2}{n_1 + n_2} + 1 
\end{equation}

Where \(n_1\) and \(n_2\) represent the occurrences of the two distinct values in the sequence.

\begin{equation}
\label{ex9}
Z = \frac{R - E(R)}{\sqrt{V(R)}} 
\end{equation}

With \(V(R)\) denoting the variance of the number of runs, expressed as the formula \ref{ex10}:

\begin{equation}
\label{ex10}
 V(R) = \frac{2n_1n_2(2n_1n_2 - n_1 - n_2)}{(n_1 + n_2)^2(n_1 + n_2 - 1)}
\end{equation}

The $p$-value, a measure in hypothesis testing, is calculated based on the standard normal distribution. For a two-tailed test, the $p$-value is obtained by finding the observing probability of a test statistic as intense as the one calculated, both in the left and the right tail of the standard normal distribution \cite{Gibbons, Rukhin, Karras}.

\begin{equation}
\label{ex11}
\text{p-value} = 2 \times P(Z > |Z|)
\end{equation}

The p-value, obtained from the standard normal distribution, determines whether to reject the null hypothesis. A higher absolute value of the test statistic signifies stronger evidence against the null hypothesis. The decision to reject the null hypothesis involves comparing the p-value to the significance level (\(\alpha\)), which affects the balance between Type I and Type II errors in hypothesis testing \cite{Gibbons, Rukhin, Karras}.

\subsubsection{The Chi-Square Test in Randomness Assessment}

The Chi-Square \cite{Plackett} test is a statistical tool that can be used to assess randomness in sequential data analysis. It quantifies the disparity between observed and expected frequencies of discrete events within a sequence. The test involves counting occurrences of each distinct element and computing the Chi-Square statistic (\( \chi^2 \)) using the formula \ref{ex12}. The statistic and associated p-value provide insights into the degree of randomness in the dataset.

\begin{equation}
\label{ex12}
\chi^2 = \sum \frac{(O_i - E_i)^2}{E_i} 
\end{equation}

Where \( O_i \) represents the observed frequency of the \( i-th \) category, and \( E_i \) represents the expected frequency of the \( i -th\) category.

The test evaluates two hypotheses: the Null Hypothesis ($H_0$) assumes observed distribution aligns with expected, indicating no significant deviation from randomness, while the Alternative Hypothesis ($H_1$) suggests otherwise. A lower p-value indicates a higher likelihood of rejecting the null hypothesis. The Chi-Square function in the `scipy` \cite{scipy} library computes the chi-squared statistic and its associated p-value. The statistic quantifies the disparity between observed and expected counts, while the p-value indicates the probability of such divergence under the null hypothesis assumption. Lower p-values suggest noteworthy deviations from the expected uniform distribution. The function allows adjustments for degrees of freedom and handling axes, enhancing adaptability to different testing scenarios.

\subsubsection{Auto-correlation}

In time-series analysis, the Auto-correlation Function (ACF) assesses the correlation between a series and its past values. The `statsmodels` \cite{statsmodels} in Python computes ACF. The ACF \cite{Stocal} at lag \(k\) is computed for a given time series \(x = \{x_1, x_2, \ldots, x_n\}\) with \(n\) observations. The auto-correlation coefficient (\(\rho_k\)) at lag \(k\) is determined using the formula \ref{ex13} :

\begin{equation}
\label{ex13}
\rho_k = \frac{\sum_{t=k+1}^{n}(x_t - \bar{x})(x_{t-k} - \bar{x})}{\sum_{t=1}^{n}(x_t - \bar{x})^2} 
\end{equation}

Here, \(\rho_k\) represents the auto-correlation coefficient, \(x_t\) denotes the observation at time \(t\), and \(\bar{x}\) is the mean of the time series. This formula \ref{ex13} elucidates the correlation between time \(t\) and \(t-k\) observations for a given lag \(k\) \cite{Stocal}. In ACF interpretation, \(\rho_k = 1\) indicates perfect positive auto-correlation, \(\rho_k = 0\) denotes no auto-correlation, and \(\rho_k = -1\) signifies perfect negative auto-correlation \cite{Stocal}.

\subsubsection{Entropy Testing}

Entropy testing constitutes a method for assessing the randomness and uncertainty inherent in sequences of random numbers, offering a quantitative measure of unpredictability. In Python, the `entropy` function within the Stats module from the `scipy` \cite{scipy} provides two essential entropy metrics, each defined by the formula \ref{ex14}. Shannon Entropy measures the expected information content within a probability distribution. The formula for Shannon Entropy $(H)$ is given by \cite{entropy1,entropy2}:

\begin{equation}
\label{ex14}
H(X) = - \sum_{i=1}^{n} P(x_i) \cdot \log_2(P(x_i))
\end{equation}

Where \(n\) is the number of distinct values in the distribution, and \(P(x_i)\) is the probability of occurrence of the \(i^{th}\) value. In the context of random number series, a higher Shannon Entropy value indicates increased unpredictability and complexity, while a lower value suggests a more ordered and less intricate sequence.

\subsubsection{Spearman Rank-Order Test}

The Spearman rank-order correlation coefficient  \cite{Zwillinger} is a non-parametric measure used to assess associations between variables based on the ranking of their observations. Computed using the test in Python's `scipy.stats` module \cite{scipy}, the correlation coefficient (\( \rho \)) ranges from -1 to +1. A value of 0 implies no correlation, while -1 or +1 indicates a perfect monotonic relationship between variables, with +1 denoting a positive correlation and -1 a negative correlation. The formula for \( \rho \) is given by expression \ref{ex15}, where \(d_i\) represents disparities between ranks of corresponding pairs and \(n\) is the total number of observations.

\begin{equation}
\label{ex15}
\rho = 1 - \frac{6 \sum d_i^2}{n \cdot (n^2 - 1)}
\end{equation}

 A smaller p-value suggests strong evidence against the null hypothesis of no correlation. The null hypothesis is rejected if \(p < \alpha\), the significance level, indicating a statistically significant monotonic relationship \cite{Zwillinger}. Conversely, a \(p > \alpha\) suggests insufficient evidence for a meaningful association \cite{Zwillinger}. Therefore, considering the p-value aids robust conclusions in statistical analyses. This research utilized the Spearman test to assess the correlation relationships between the training data and the model's generated data.

\subsubsection{Kendall's Tau  Test}

Kendall's Tau is a non-parametric correlation coefficient used to measure the strength and direction of the monotonic relationship between two variables. The calculation involves analyzing concordant and discordant pairs of observations, where pairs are considered concordant if their relative rankings align and discordant if they have opposite orders in their rankings. The Kendall's score (\(S_t\)) is computed as the sum of concordant and discordant pairs, as per Equation \ref{ex16}, and is subsequently normalized to fall within [-1, 1] range, as shown in Equation \ref{ex17}. A \( \tau \) value of 1 indicates perfect favorable agreement, -1 signifies perfect negative agreement, and 0 suggests no correlation between the ranked sets \cite{Zwillinger}.

\begin{equation}
\label{ex16}
\sum_{i=1}^{n-1} \sum_{j=i+1}^{n} \text{sign}(\text{rank}_a[j] - \text{rank}_a[i]) \times \text{sign}(\text{rank}_b[j] - \text{rank}_b[i]) 
\end{equation}

\(S_t\) can reach a maximum of \(\frac{n(n-1)}{2}\) when the two rankings are parametrical and a minimum of \(-\frac{n(n-1)}{2}\) when the sets are ranked in the opposite order. Kendall's Tau is subsequently normalized to fall within the range [-1, 1], represented by the formula \ref{ex17}:

\begin{equation}
\label{ex17}
\tau = \frac{S_t}{\frac{n(n-1)}{2}}
\end{equation}

This normalization facilitates the interpretation of \(\tau\), where  1 denotes perfect favorable agreement, -1 signifies perfect negative agreement, and 0 indicates no correlation between the ranked sets \cite{Zwillinger}.

The `scipy` library  \cite{scipy} provides this test  to compute Kendall's Tau and the associated p-value for assessing statistical significance. Positive \( \tau \) denotes a positive monotonic relationship, while negative \( \tau \) indicates an inverse relationship. A value near zero suggests no monotonic relationship. The associated p-value helps determine statistical significance, with a small value indicating significance. If the calculated p-value is less than the chosen significance level  $\alpha $, the null hypothesis of no correlation is rejected, suggesting a statistically significant monotonic relationship between the variables. Conversely, a p-value greater than $\alpha $  suggests insufficient evidence to assert a substantial association.

\subsubsection{Spectral  Test}

The Fourier Transform is a mathematical tool that converts a function that varies over time (or space) into its representation in the frequency domain. Data point that is measured at discrete points in time, the Power Spectral Density (PSD) is commonly estimated using the Discrete Fourier Transform (DFT) \cite{Dempster, Oppenheim}. 
In this context, a number points like as signal that varies over discrete time, denoted by ( x[n] ), where ( n ) represents the index of the discrete time. The DFT of this signal can be expressed as per the formula in Equation \ref{ex18} \cite{Dempster, Oppenheim}:

The Fourier Transform is a mathematical tool that transforms a function varying over time into its representation in the frequency domain. For discrete-time signals represented by \( x[n] \), the Discrete Fourier Transform (DFT) is commonly employed to estimate the Power Spectral Density (PSD) \cite{Dempster, Oppenheim}. The DFT of a signal \( x[n] \) is given by Equation \ref{ex18}, where \( N \) is the total number of samples, and \( k \) is the frequency index.  \( \omega_k = \frac{2\pi}{N}k \) gives the corresponding angular frequency \( \omega_k \).

\begin{equation}
\label{ex18}
 X[k] = \sum_{n=0}^{N-1} x[n] e^{-i\frac{2\pi}{N}kn}
\end{equation}

The PSD estimate \( S_x(\omega_k) \) is obtained by squaring the magnitude of the DFT and normalizing by \( N \) (Equation \ref{ex19}). Alternatively, in terms of frequency index \( k \), the PSD can be expressed as \( S_x(f_k) \) (Equation \ref{ex20}), where \( f_k \) represents the frequency in Hertz corresponding to index \( k \) \cite{Dempster, Oppenheim}.

\begin{equation}
\label{ex19}
S_x(\omega_k) = \frac{1}{N} \left| X[k] \right|^2 
\end{equation}

Alternatively, in terms of frequency index \( k \), the PSD can be expressed as The Equation \ref{ex20}:

\begin{equation}
\label{ex20}
S_x(f_k) = \frac{1}{N} \left| X[k] \right|^2
\end{equation}

Methods like the periodogram or Welch method are often employed to improve the accuracy of PSD estimation from a finite-length signal. These methods involve segmenting the signal, computing PSD for each segment, and averaging them \cite{Welch}. The PSD reveals power distribution across different frequencies, aiding in identifying dominant frequency components and analyzing frequency content. 

\subsubsection{Kolmogorov-Smirnov (KS) Test}

The Kolmogorov-Smirnov (KS) test, implemented by `scipy.stats.kstest` \cite{scipy}, is a  nonparametric statistical tool utilized to assess the goodness of fit of various probability distributions \cite{Fasano}. It evaluates whether a dataset conforms to the parameters of a specified distribution, such as the Gamma distribution, under the null hypothesis $(H\_0)$.

In practical terms, the KS test compares the empirical cumulative distribution function (CDF) derived from the sample data with the theoretical CDF of the hypothesized distribution. The maximum vertical distance between the two CDFs yields the KS statistic, indicating the discrepancy. Acceptance or rejection of the null hypothesis is based on comparing this statistic with critical values.

The KS test assesses the adherence of generated data to specific probability distributions, including normal, uniform, exponential, binomial, Poisson, gamma, Weibull with minimum bound, and Weibull with maximum bound distributions. 

\subsubsection{Statistical Test Suite for Random Number Generators}

To evaluate the statistical quality of the generated random numbers, we employed a comprehensive test suite comprising statistical tests, derived and adapted from the NIST SP 800-22 standard \cite{Rukhin}. These tests aim to measure various aspects of randomness and entropy, particularly in bit sequences. 
A distinguishing feature of our implementation is the ability to process not only binary files, as in the original NIST SP 800-22 suite, but also sequences composed of floating-point numbers. For this purpose, we developed a specialized auxiliary function, which transforms floating-point numbers into binary sequences using a configurable thresholding or conversion mechanism. Furthermore, the system is capable of batch-processing datasets stored in pickle format, enabling efficient large-scale evaluation of random number generation models.

\subsubsection{Forward Secrecy and Next-Bit Unpredictability Evaluation}
\label{sec:fs-nb}

To assess the cryptographic robustness of the AI-Hybrid TRNG, we employed the dual-test methodology proposed in \cite{AN2022}, combining \emph{Hamming Distance} (HD) analysis with a \emph{Practical Next-Bit Prediction} (PNB) test. The HD test evaluates forward secrecy, while the PNB test measures next-bit unpredictability.

\paragraph{Test Block Construction.}
Each 2048-bit output block from the generator is partitioned into two equal halves: $1024$ bits labeled as \textit{leak} (assumed publicly observable) and $1024$ bits labeled as \textit{true} (considered secret). In all subsequent tests, $m = 1024$ denotes the number of bits used per statistical decision.

\paragraph{Hamming Distance Test (Forward Secrecy).}
This test quantifies the bit-wise difference between the \textit{true} portions of two consecutive output blocks (before and after reseeding), i.e., the Hamming distance $\mathrm{HD}$.
Under the assumption of independent and uniformly random bit generation, the expected distribution is $\mathrm{HD} \sim \mathrm{Bin}(m, 0.5)$.
A two-sided binomial test is applied at significance level $\alpha = 0.01$, flagging any deviation from this expected behavior as statistically significant evidence against independence, potentially indicating compromised forward secrecy.

\paragraph{Practical Next-Bit Prediction Test (PNB).}
To evaluate next-bit unpredictability, we simulate an adversarial model that attempts to predict each bit in the \textit{true} half using preceding bits from the \textit{leak} half. Specifically, a sliding window of $K = 16$ consecutive bits is extracted from the \textit{leak} portion to train a logistic regression model. This model is then tested on the remaining $20\%$ of the data to estimate prediction accuracy over $m = 1024$ trials per block.

Under ideal randomness, no predictive model should perform significantly better than random guessing. For binary classification, the expected accuracy under pure chance lies within $[0.40, 0.60]$ with confidence $\alpha = 0.01$. Thus, any block where the prediction accuracy exceeds $0.60$ is flagged as a \emph{failure}, suggesting potential predictability in the TRNG output.

\paragraph{Summary.}
The results from both the Hamming Distance and Practical Next-Bit Prediction tests indicate that the AI-Hybrid TRNG maintains strong cryptographic properties: reseeding operations do not introduce detectable correlations, and no meaningful advantage is gained by an adversary attempting to predict future outputs from past observations. These findings support the conclusion that the generator upholds both forward secrecy and next-bit unpredictability under practical adversarial assumptions.

\begin{figure*}[!t]
\centering
\includegraphics[width=\textwidth,height=0.8\textheight,keepaspectratio]{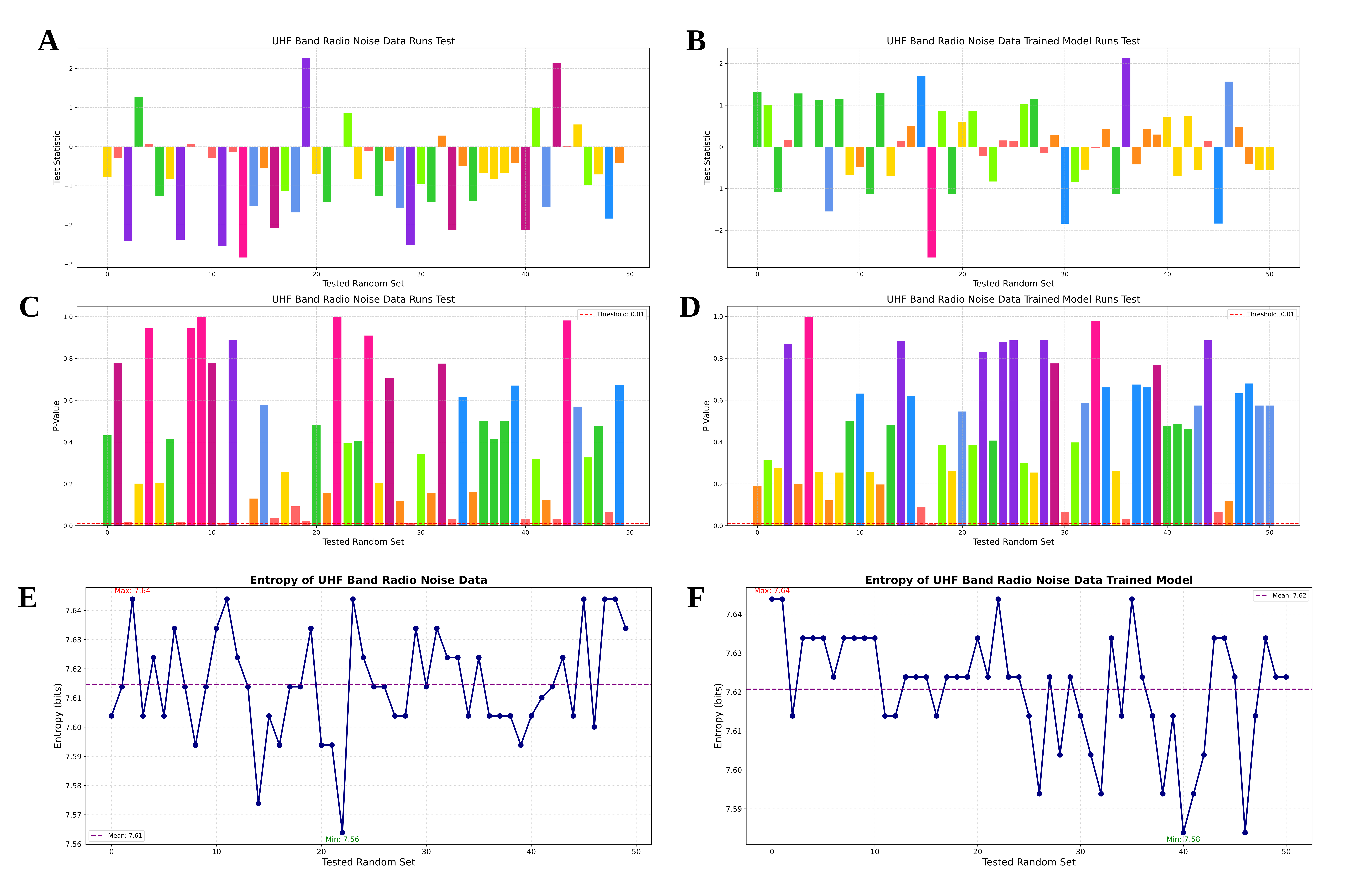}
\caption{\textbf{The analysis compares the proposed AI model with UHF Band Radio Noise Data using Runs Test and Entropy metrics. } Each bar represents different random sets with colors indicating magnitude. The Runs Test Statistic for the noise data (A) and AI model (B) show similar patterns: positive z-values indicate frequent alternation (high mixing), while negative z-values suggest clustering. P-values for the noise data (C) and AI model (D) indicate randomness probability, with low p-values (<0.01) suggesting non-randomness. Entropy values are shown for the noise data (E) and AI model (F), demonstrating the AI model's superior performance based on its training data (pre-deployment
evaluation). }
\label{fig:re3}
\end{figure*}

\begin{figure*}[!t]
\centering
\includegraphics[width=\textwidth]{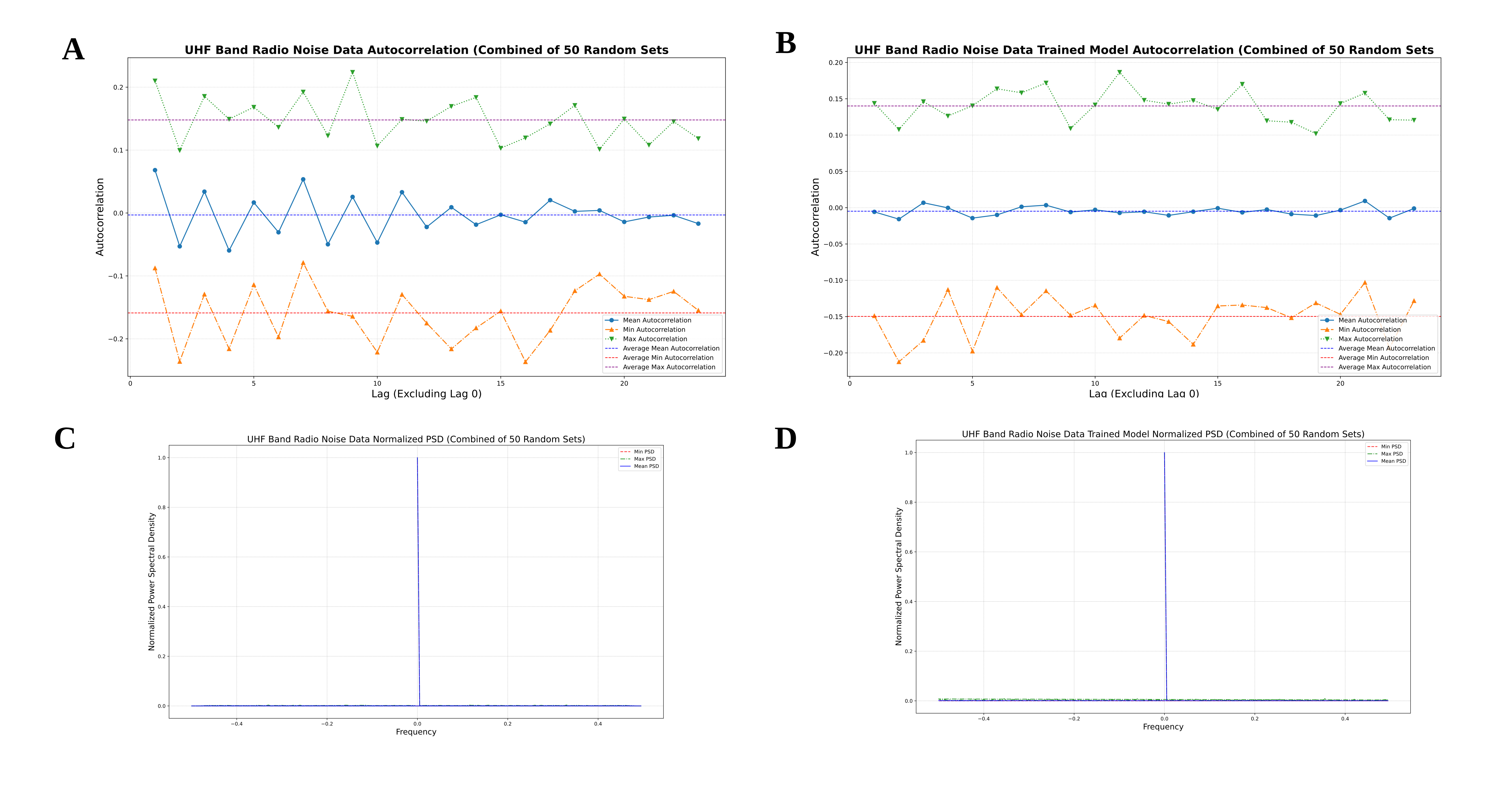}
\caption{\textbf{Autocorrelation and Normalized Power Spectral Density Analysis: Proposed AI Model vs. UHF Band Radio Noise Data} A) Displays the autocorrelation of UHF Band Radio Noise Data across different lags for 50 combined random sets. (B) Shows the autocorrelation results for the proposed AI model, mirroring the noise data. In both (A) and (B), the blue line represents the mean autocorrelation across random sets, the green line shows the maximum autocorrelation at each lag, the orange line indicates the minimum autocorrelation at each lag, the dotted line marks the average mean autocorrelation, and the dashed line indicates the average maximum autocorrelations. (C) Presents the Normalized Power Spectral Density (PSD) of the UHF Band Radio Noise Data, depicting how the signal's power is distributed over different frequencies. (D) shows the normalized PSD for the proposed AI model, which is consistent with the noise data. The blue lines in (C) and (D) represent the mean normalized PSD. The proposed AI model demonstrates superior performance based on its training data (pre-deployment
evaluation).}
\label{fig:re2}
\end{figure*}

\begin{figure*}[!t]
\centering
\includegraphics[width=\textwidth]{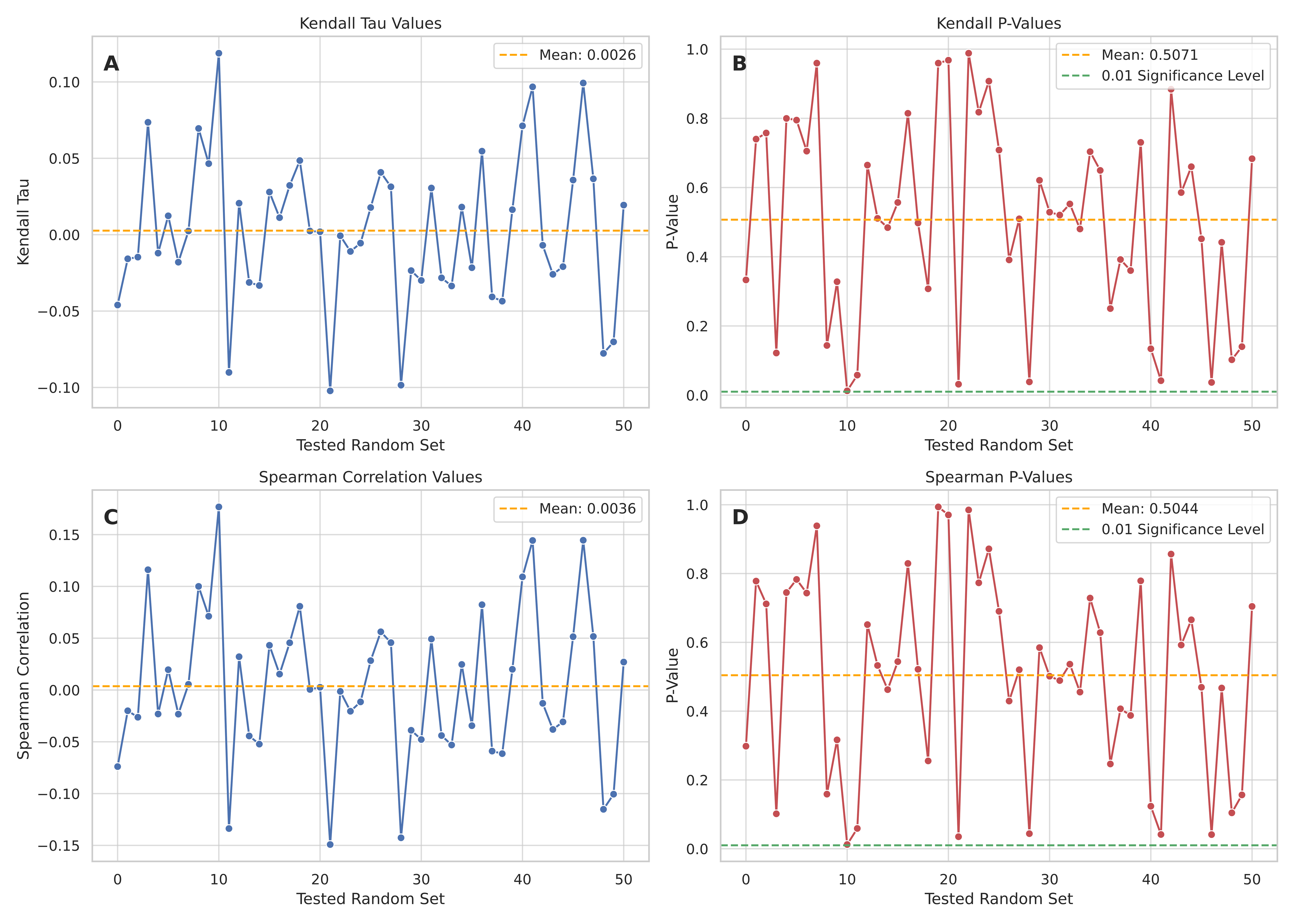}
\caption{\textbf{Correlation and Significance Analysis Using Kendall Tau and Spearman Correlation} The correlation and significance analysis using Kendall Tau and Spearman correlation under the same parameter settings in the proposed AI model reveals the following: (A) Kendall Tau correlation values for 51 tested random sets fluctuate around zero, with a mean value of 0.0026 (dashed orange line), indicating no significant correlation trend between the generated and trained data. (B) The p-values for the Kendall Tau correlation suggest the statistical significance of these observations. (C) Spearman correlation values for the same 51 random sets measure rank correlation, while (D) the associated p-values indicate the significance of these correlations. Overall, the proposed AI model shows no significant correlation trend, as indicated by the low mean Kendall Tau value and corresponding p-values (pre-deployment
evaluation).}
\label{fig:re1}
\end{figure*}

\begin{table*}[!t]
  \centering
  \footnotesize
  \setlength\tabcolsep{4pt}
  \renewcommand{\arraystretch}{0.9}
  \caption{
  Pass Rates (\%) of Statistical Tests on 50 Sets of CPU-Generated and Radio-Based Noise Data and Corresponding AI-Generated Outputs (pre-deployment evaluation)
}
  \label{table:pass-rates}
  \begin{tabular}{lcccc}
    \toprule
    \textbf{Test Type} 
      & \textbf{CPU-Based} 
      & \textbf{CPU-Based AI} 
      & \textbf{Radio-Based} 
      & \textbf{Radio-Based AI} \\
    \midrule
    ACF Mean (Min, Mean, Max) &
      	-0.11, 0.03, 0.20 &
      	-0.15, -0.01, 0.14 &
      	-0.16, 0, 0.15 &
      	-0.15, 0, 0.14 \\
    Augmented Dickey--Fuller Test
      & 100\% & 100\% & 92\%  & 100\% \\
    ADF-GLS Test
      & 100\% & 100\% & 92\%  & 100\% \\
    Phillips--Perron Test
      & 100\% & 100\% & 100\% & 100\% \\
    Entropy (Min, Mean, Max) 
      & 6.13, 6.25, 6.36 
      & 5.64, 5.94, 6.24 
      & 7.56, 7.61, 7.64 
      & 7.58, 7.62, 7.64 \\
    Chi-Square Test 
      & 100\% & 98\%  & 90\%  & 100\% \\
    Durbin--Watson Test 
      & 100\% & 100\% & 100\% & 100\% \\
    Run Test 
      & 98\%  & 100\% & 98\%  & 100\% \\
    Spectral Analysis 
      & 100\% & 100\% & 100\% & 100\% \\
    Weibull Min Test 
      & 100\% & 100\% & 100\% & 100\% \\
    Gamma Distribution Test 
      & 100\% & 100\% & 100\% & 100\% \\
    Logistic Distribution Test 
      & 100\% & 100\% & 100\% & 100\% \\
    Poisson Distribution Test 
      & 100\% & 100\% & 100\% & 100\% \\
    KS Normality Test 
      & 100\% & 100\% & 100\% & 100\% \\
    KS Uniformity Test 
      & 100\% & 100\% & 100\% & 100\% \\
    KS Exponential Test 
      & 100\% & 100\% & 100\% & 100\% \\
    KS Binomial Test 
      & 100\% & 100\% & 100\% & 100\% \\
    Spearman Test 
      & 100\% & 100\% & 98\%  & 98\%  \\
    Kendall’s Tau Test 
      & 100\% & 100\% & 98\%  & 98\%  \\
    \bottomrule
  \end{tabular}
\end{table*}

   %


\end{document}